\documentclass[%
 aip,
 pre,%
 amsmath,amssymb,
reprint,%
]{revtex4-1}
\usepackage {CJK}
\usepackage{graphicx}% Include figure files
\usepackage{dcolumn}% Align table columns on decimal point
\usepackage{bm}% bold math
\usepackage[mathlines]{lineno}% Enable numbering of text and display math
\usepackage{color}
\usepackage{gensymb}
\usepackage{hyperref}% add hypertext capabilities
\begin{document}
\begin{CJK*}{GB}{gbsn}
%\begin {CJK}{GB} {song} % Use default fonts from CJK (see below)
\preprint{AIP/123-QED}

\title{Cavity deformation and bubble entrapment during the impact of droplets on a liquid pool}
\author{Zhigang Xu}
\affiliation{State Key Laboratory of Engines, Tianjin University, Tianjin, 300072, China.}%
\author{Tianyou Wang}
\affiliation{State Key Laboratory of Engines, Tianjin University, Tianjin, 300072, China.}%
\author{Zhizhao Che}
\email{chezhizhao@tju.edu.cn}
 \affiliation{State Key Laboratory of Engines, Tianjin University, Tianjin, 300072, China.}

\date{\today}% It is always \today, today,

\begin{abstract}
The impact of droplets on a liquid pool is ubiquitous in nature and important in many industrial applications. A droplet impacting on a liquid pool can result in the pinch-off of a regular bubble or entrap a large bubble under certain impact conditions. In this study, the cavity deformation and the bubble entrapment during the impact of droplets on a liquid pool are studied by combined experimental measurements and numerical simulations. The time evolution of the free surface profile obtained in the numerical simulation is in good agreement with the experimental results. The cavity created by the droplet impact affects the pinch-off of regular bubbles and the entrapment of large bubbles. The regular bubble pinch-off is the direct consequence of the capillary wave propagating downward along the interface of the cavity and merging at the bottom of the cavity. In contrast, the large bubble entrapment is due to the merging of the liquid crowns at the mouth of the cavity. Gravity and environmental pressure play important roles in cavity deformation and bubble entrapment after droplet impact on liquid pools. The maximum depth of the cavity decreases as the gravitational effect becomes stronger. The phenomenon of regular bubble pinch-off may disappear as the gravity decreases. We find that the regular bubble pinch-off can transform into large bubble entrapment when increasing environmental pressure. The size of the large bubble entrapped decreases with increasing the environmental pressure due to the larger difference in the pressure and the vorticity around the liquid crown at increased environmental pressure. Finally, the regime map of bubble entrapment after the droplet impact is obtained.
\end{abstract}

%\pacs{47.55.D-, 47.55.N-, 47.61.Jd, 47.32.C-}% PACS, the Physics and Astronomy
%47.55.D-	Drops and bubbles
%47.55.N-	Interfacial flows
%47.61.Jd	Multiphase flows
%47.32.C-	Vortex dynamics

%\keywords{Compound droplet; Droplet; Vortex; Multiphase microfluidics; microchannel}%Use showkeys class option if keyword
\maketitle
\end{CJK*}
%\tableofcontents

\section{Introduction}\label{sec:sec1}

Droplet impact on liquid surfaces occurs in many natural and industrial processes, such as precipitation \cite{Raes2000CyclingAerosol}, spray coating \cite{Guo2016CrownFormation}, fuel atomization \cite{Moreira2010DropletImpactICE}, and spray cooling \cite{Horacek2005NozzleSpray, Kim2004CoolingPlain, Shedd2005SprayCooling, Xie2012FilmFlow}. The outcomes of the impact mainly include bouncing \cite{Tang2019InterfacialGas, Wang2008LowVelocity, Beesabathuni2015ImmiscibleInterface}, coalescence \cite{Blanchette2009, Geri2017DelayCoalescence, Kulkarni2021CoalescenceSpreading}, jet breakup \cite{Agbaglah2015VortexJet, Liu2018WindDropJet, Pan2010SurfaceProperties, Blanco2021JetBurstingBubbles, Das2022EvolutionJetPool}, and splashing \cite{Lee2015VortexSplashing, Leng2001SplashSphere, Motzkus2011AirborneParticle}. In addition, the impact can result in the entrapment of air bubbles \cite{Deng2007ViscositySurface, Oguz1990BubbleEntrainment, Haldar2018ChemicallyGelation}, which is often accompanied by high-speed jets \cite{Rein1996TransitionalRegime}. The bubble entrapment is known to be important for the generation of noise, the production of aerosols, and the transport of gases across the ocean surface \cite{Deane2002ScaleBubble, Franz1959SplashSound, Pumphrey1990EntraimentBubble}.

The pinch-off of bubbles after the droplet impact on liquid pools has been considered in many studies \cite{Franz1959SplashSound, Pumphrey1988SoundEmittedWater, Chen2014ViscosityEffect, Ray2012BubbleScaling, Morton2000FlowRegime, Ray2015RegimesCoalescence, Oguz1990BubbleEntrainment, Pumphrey1990EntraimentBubble, Deng2009BubblePhenomenon, Deng2007ViscositySurface}. The impact creates a cavity at the impact point, and then a single bubble is pinched off at the cavity bottom as the cavity contracts, and this bubble was named a \emph{regular bubble}. The Weber number $\text{We}\equiv \rho U_{0}^{2}{{D}_{0}}/\sigma $ can be used to indicate the ratio between inertial and capillary forces, where $\rho $ and $\sigma $ are the density and surface tension coefficient of the fluid, ${{U}_{0}}$ and ${{D}_{0}}$ are the speed and diameter of the droplet, respectively. Bay et al. \cite{Ray2015RegimesCoalescence} studied the effect of the inertial force on the crater shape just before collapse, and found that the crater width and wave swell angle (i.e., the angle of the crater cusp just before the crater collapses) increase as the We increases and the regular bubble pinch-off occurs for wave swell angles ranging from 100${}^\circ $ to 120${}^\circ $. Capillary waves were found to travel along the interface and merge at the cavity bottom to produce a regular bubble \cite{Deng2009BubblePhenomenon, Leng2001SplashSphere, Rein1996TransitionalRegime}.  The capillary number $\text{Ca}\equiv \mu {{U}_{0}}/\sigma $ can be used to indicate the ratio between viscous and capillary forces, where $\mu $ is the fluid dynamic viscosity. The viscosity and the surface tension could affect the regular bubble pinch-off, and the regular bubble pinch-off cannot be found when the capillary number is larger than 0.6 \cite{Deng2007ViscositySurface}. The regular bubble pinch-off was also studied numerically using the volume of fluid (VOF) method \cite{Berberovic2011InertiaFlow, Morton2000FlowRegime, Oguz1990BubbleEntrainment, Ray2012BubbleScaling}. Oguz et al. \cite{Oguz1990BubbleEntrainment} found that whether a bubble is entrapped or not is determined by the competition between the timescale for the cavity to achieve its maximum depth and the timescale of the capillary wave. Morton et al. \cite{Morton2000FlowRegime} described the importance of the capillary wave propagation to the bubble pinch-off, and vortex rings were found in the process of the bubble pinch-off. The entrapped bubble was found to depend on the cavity shape and its maximum depth \cite{Ray2012BubbleScaling}. Chen et al. found \cite{Chen2014ViscosityEffect} that both the lower and upper limits of the regular bubble entrapment region increased with the liquid viscosity, and scaling models for the lower and upper limits of regular bubble entrapment were proposed.

A \emph{large bubble} can be entrapped after the closure of the crown during the droplet impact on a pool at large Weber numbers \cite{Xu2018TemperatureBubble, Bisighini2010CraterEvolution}. The size of the entrapped large bubble can be several times larger than the initial droplet \cite{Zou2012LargeBubble}. Numerical and experimental studies on the cavity dynamics \cite{Dake2017BubbleVortex} found that large bubbles were entrapped during the impact of prolate-shaped droplets. Large bubble entrapment was also found to be a vortex-driven phenomenon after the impact of an oscillating droplet \cite{Thoraval2016VortexRing}.

The deformation of the cavity and the entrapment of bubbles result from the interplay of inertia, viscous force, surface tension, and gravity. Most existing studies focus on the effects of inertial force, viscous force, and surface tension force \cite{Deng2007ViscositySurface, Deng2009BubblePhenomenon, Ray2015RegimesCoalescence, Zeff2000SingularityDynamic}. However, there are few studies on the effect of gravity, even though the gravitational force can significantly alter the evolution of the interface and affect bubble entrapment. The impact of droplets at high environmental pressure is crucial to many relevant applications, such as fuel atomization in internal combustion engines \cite{Panao2005SprayInjectionSystem, Singh2007EngineCombustionModel} and the droplet impact in nuclear power plants \cite{Zhang2020DropletNuclearInstallation, Kanazawa1995CoolantDropletBehavior, Okawa2007DropletsAnnularFlow}. In addition, the impact of droplets at low gravity is also important to many relevant applications, such as spray cooling in moving systems and fire extinguishing in spacecraft \cite{Tao2013DropletSolidMicrogravity, Han2014WaterMistMicrogravity}. The effect of the environmental pressure has been proved to be very important during the impact of droplets on solid surfaces \cite{Xu2007SplashInterplay, Xu2005Splashing}. In addition, the effect of the environmental pressure was also important in the droplet impact on a liquid pool \cite{Klyuzhin2010PersistingDropletsSurfaces, Marcotte2019EjectaCorollaSplashes}. Klyuzhin et al. \cite{Klyuzhin2010PersistingDropletsSurfaces} studied the effect of the environmental pressure on the droplet floating time when droplets impact a liquid pool at a low impact speed, and they found that the floating time decreases linearly with decreasing the environmental pressure. In addition, Marcotte et al. \cite{Marcotte2019EjectaCorollaSplashes} studied the effect of the environmental pressure on splashing when ethanol droplets impact liquid pools of different viscosities, and they found that decreasing the environmental pressure could suppress splashing for high-viscosity liquid pools but could not suppress splashing for low-viscosity liquid pools. In our study, the effect of the environmental pressure on bubble entrapment during droplet impact on liquid pools is studied.

In this paper, we focus on the cavity deformation and bubble entrapment after droplets impact a liquid pool. We aim to reveal the role of gravity and environmental pressure on regular bubble pinch-off and large bubble entrapment. The cavity shape and its depth are studied at different gravities in our simulations since the cavity shape and the timescale for the cavity to achieve its maximum depth are important to the bubble pinch-off. The regimes of bubble pinch-off and the theoretical threshold at different gravities are studied here, while previous studies \cite{Deng2007ViscositySurface, Oguz1990BubbleEntrainment, Pumphrey1990EntraimentBubble, Deng2009BubblePhenomenon, Morton2000FlowRegime, Ray2012BubbleScaling, Chen2014ViscosityEffect, Ray2015RegimesCoalescence} mainly considered earth's gravity. Inspired by previous research showing that the environmental pressure is important to the droplet impact on liquid pools \cite{Klyuzhin2010PersistingDropletsSurfaces, Marcotte2019EjectaCorollaSplashes}, we study the effect of the environmental pressure on the bubble entrapment during the impact process. The regimes of bubble entrapment at different environmental pressures are studied, and the regular bubble pinch-off is found to transform into large bubble entrapment as the environmental pressure increases, whereas the large bubble entrapment was previously observed mainly at large Weber numbers at standard environmental pressure \cite{Xu2018TemperatureBubble, Bisighini2010CraterEvolution}. The experimental setup is introduced in Sec.\ \ref{sec:sec2}. The numerical methodology is presented in Sec.\ \ref{sec:sec3}. Results are discussed and analyzed in Sec.\ \ref{sec:sec4}, including the model validation, the effects of gravity and environmental pressure on the cavity deformation and bubble entrapment. Conclusions are summarized in Sec.\ \ref{sec:sec5}.

\section{Experimental setup}\label{sec:sec2}

\begin{figure}
  \centering
  \includegraphics[width=\columnwidth]{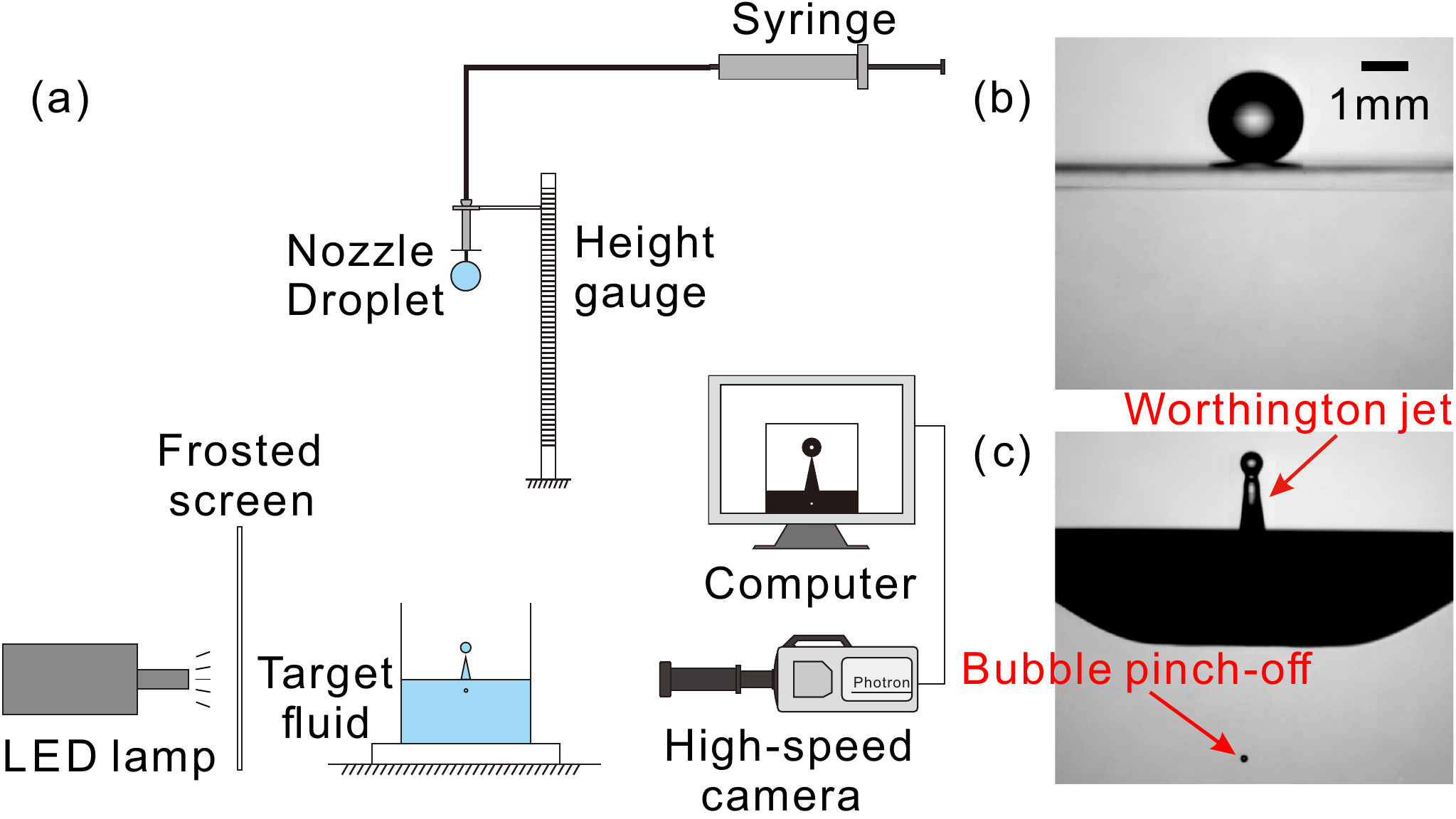}\\
  \caption{(a) Schematic diagram of the experimental setup. (b) Droplet at the instant of impact. (c) Bubble pinch-off and Worthington jet after the impact.}\label{fig:fig01}
\end{figure}

The setup for the impact experiment is schematically shown in Fig.\ \ref{fig:fig01}. By pushing a syringe to produce a droplet at a syringe tip, the droplet fell and impacted the target liquid pool. The liquid pool is in an open acrylic container with a size of $40\times 40\times 30$ mm$^3$. The width of the container is more than $10 D_0$, and the depth of the container is more than 6 times the maximum depth of the cavity. The influence of the container's wall on the impact process can be neglected. A high-speed camera (Photron Fastcam SA1.1) was used to film the impact process from the side. The resolution is $1024\times 896$ pixels, corresponding to a spatial resolution of 15.9 $\mu \text{m/pixel}$. The frame rate is 6000 frames per second (fps), and the exposure time is 1/30000 seconds. An LED light was used as the light source, combined with a frosted screen to diffuse the light. A typical image at the impact instant is shown in Fig.\ \ref{fig:fig01}b, and a typical regular bubble after the impact is shown in Fig.\ \ref{fig:fig01}c. Using a customized image-processing program of Matlab, we measured the droplet size, the impact speed, and the cavity deformation from the images.

The droplet and the pool were prepared using the same liquid, a mixture of glycerin and water with a mass fraction of glycerin of 62\%. The physical properties of the mixture were measured experimentally. We measured the density of the mixture by an electronic liquid densitometer (WLD-MD01, XIONGFA), measured the surface tension of the mixture by a platinum ring tensiometer (MC-1021, MINCE), and measured the dynamic viscosity of the mixture by a digital viscometer (NDJ-8S, SHANGPU). The room temperature is 26.5 ${}^\circ $C. The density of the mixture is 1156 kg/m$^3$, the surface tension is 68 mN/m, and the dynamic viscosity is 9.5 mPa$\cdot$s.

\section{Numerical methodology}\label{sec:sec3}

The system of droplet impact consists of two immiscible and incompressible phases, i.e., the liquid and air phases. The VOF method in OpenFOAM is employed for the simulation. The OpenFOAM version used in this study is OpenFOAM-2.1.x. The continuity equation and momentum equation are respectively
\begin{equation}\label{eq:eq01}
  \nabla \cdot (\rho \mathbf{u})=0,
\end{equation}
\begin{equation}\label{eq:eq02}
  \frac{\partial (\rho \mathbf{u})}{\partial t}+\nabla \cdot (\rho \mathbf{uu})=-\nabla p+\nabla \cdot \mu [\nabla \mathbf{u}+{{(\nabla \mathbf{u})}^{\text{T}}}]+\rho \mathbf{g}+\sigma \kappa \nabla \alpha,
\end{equation}
where $\mathbf{u}$ is velocity vector, $\mu$ is dynamic viscosity, $\mathbf{g}$ is gravitational acceleration, $p$ is pressure, $\alpha $ is volume fraction, and $\kappa$ is interfacial curvature, which can be calculated as
\begin{equation}\label{eq:eq03}
  \kappa =-\nabla \cdot \left( \frac{\nabla \alpha }{\left| \nabla \alpha  \right|} \right).
\end{equation}
The interface is blurred due to numerical diffusion and the discretization of the convective term is important. To keep the sharp resolution of the interface, Weller \cite{Waller2008} add an artificial convective term into the phase fraction equation
\begin{equation}\label{eq:eq04}
  \frac{\partial \alpha }{\partial t}+\nabla \cdot (\mathbf{u}\alpha )+\nabla \cdot \left[ \alpha (1-\alpha ){{c}_{\alpha}} \left| \mathbf{u} \right| \frac{\nabla \alpha }{\left| \nabla \alpha  \right|} \right]=0.
\end{equation}
The third term on the left-hand side of Eq.\ (\ref{eq:eq04}) is an extra artificial convective term, where ${c}_{\alpha}$ is the compression coefficient and ${{c}_{\alpha }}=1$ in our study.
The physical properties can be obtained from the volume fraction
\begin{equation}\label{eq:eq05}
  \rho =\alpha {{\rho }_{l}}+(1-\alpha ){{\rho }_{g}},
\end{equation}
\begin{equation}\label{eq:eq06}
  \mu =\alpha {{\mu }_{l}}+(1-\alpha ){{\mu }_{g}},
\end{equation}
where the subscripts $l$ and $g$ refer to the properties for liquid and gas, respectively. The interFoam solver, which has been used in many previous studies \cite{Chen2014ViscosityEffect, Gupta2020SplashingMovingInterface, Hoang2013DynamicsBreakupJunction} to simulate incompressible multiphase flows, is employed in this study. Governing equations (1)-(3) are discretized using the finite volume method. Discretized linear matrices for the mass and momentum equations are solved using the Preconditioned BiConjugate Gradient (PBiCG) method and the Preconditioned Conjugate Gradient (PCG) method, respectively. A critical work for the volume fraction equation is to ensure the boundedness of the volume fraction. The multi-dimensional universal limiter with explicit solution (MULES) \cite{Damian2013ModelSimultaneousInterfaces} developed from the flux correct transport (FCT) algorithm \cite{Boris1973FluxCorrectedTransport} is used to ensure the boundedness of the volume fraction. % The interFoam solver
%${{\rho }_{g}}$ and ${{\mu }_{g}}$ are the density and the viscosity for the gas, and ${{\rho }_{l}}$ and ${{\mu }_{l}}$ for the liquid.

Since the process of cavity deformation and bubble entrapment is symmetric, axisymmetric simulations are performed by setting a wedge domain in cylindrical coordinate. The axial section of the computational domain is rectangular (20 mm $\times $ 15 mm), and the depth of the pool is 10 mm. The droplet is initialized above the pool surface with a separation of $0.1D_0$. We set the bottom of the domain as a wall with a no-slip boundary, and the side and top as open boundaries. The initial speed of the droplet is $U_0$. In the simulations of environmental pressure variation, the gas density is varied as the influence of pressure on the liquid density, liquid viscosity, and air viscosity is negligible. The properties of the air at different environmental pressures are obtained from NIST REFPROP, and the properties of the liquid are the same as that in experiments at 26.5 ${}^\circ $C. The Weber number, as a ratio between the inertia and surface tension force, is varied by varying the droplet speed ${{U}_{0}}$ from 2.1 to 4 m/s. The Bond number $\text{Bo}\equiv {\rho gD_{0}^{2}}/{\sigma }$, as a ratio between the gravitational and surface tension forces, is varied by varying the gravitational acceleration \emph{g} from 0 to 10 $\text{m}/{{\text{s}}^{2}}$. The Weber number is in a range of $157.4<\text{We}<571.2$ and the Bond number is in a range of $0<\text{Bo}<0.8$, and the corresponding Froude number $\text{Fr}\equiv {{{U}_{0}}}/{\sqrt{g{{D}_{0}}}}$, a ratio between the inertia and gravitational force, is in a range of $0<\text{Fr}<4500$.

\begin{figure*}
  \centering
  \includegraphics[width=1.4\columnwidth]{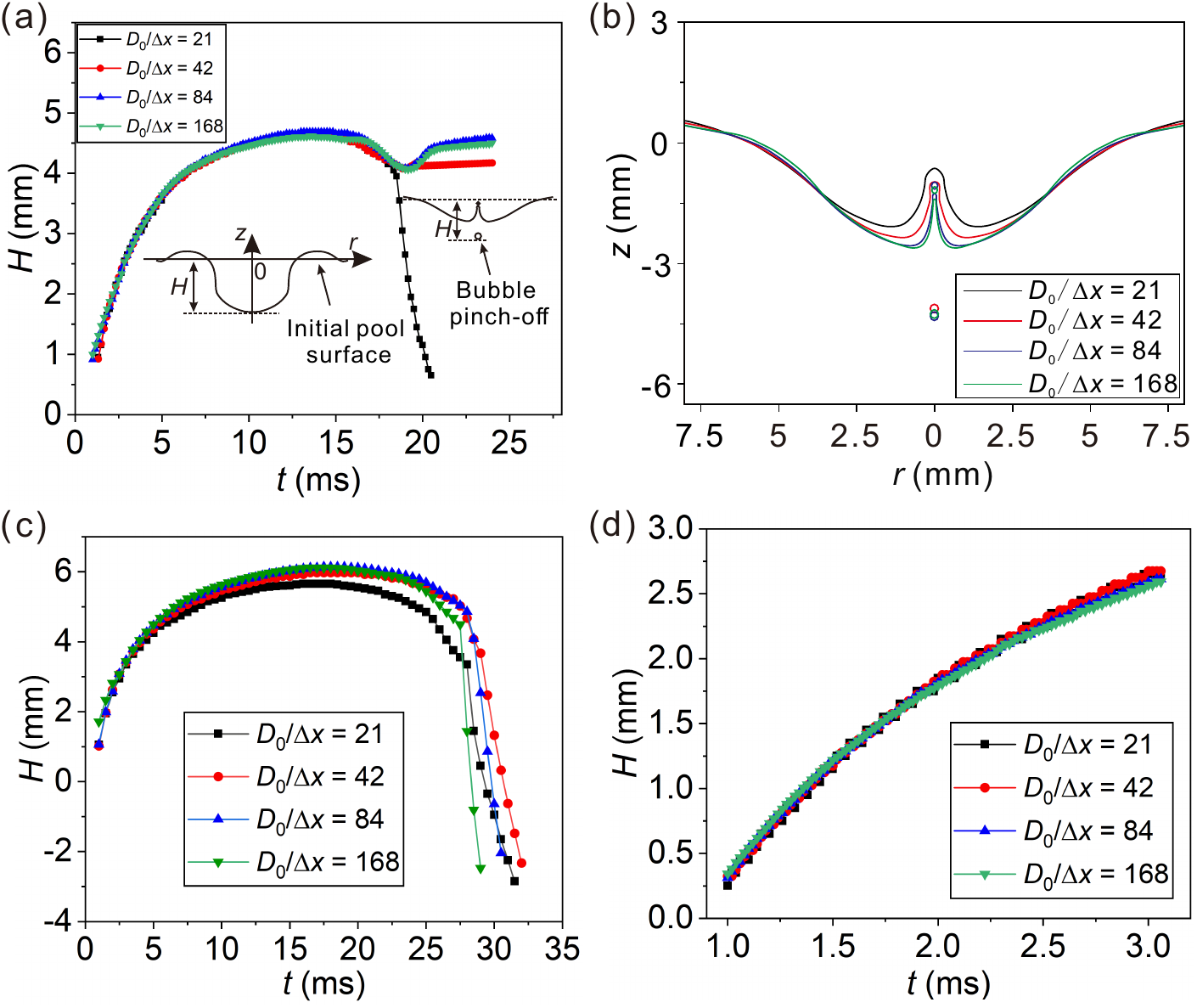}\\
  \caption{Grid-independence study for the droplet impact process. The mesh size was progressively refined from ${{D}_{0}}/\Delta x=21$ to 168. ${{D}_{0}}=2.07$ mm. (a) Cavity depth and the distance from the pool surface to the bottom of the regular bubble at $\text{We}=227$ and $P^*=1$, (b) Gas-liquid interface profiles at 20.5 ms at $\text{We}=227$ and $P^*=1$, (c) Cavity depth at $\text{We}=571$ and $P^*=1$, (d) Cavity depth during the large bubble entrapment at $\text{We}=279.9$ and $P^*=30$.}\label{fig:fig02}
\end{figure*}

A grid-independence study for the impact process was performed, as shown in Fig.\ \ref{fig:fig02}. Before the bubble pinch-off, the depth of the cavity $H$ was measured from the initial surface of the liquid pool to the bottom of the cavity, and after the bubble pinch-off, $H$ was measured from the initial surface of the liquid pool to the bottom of the regular bubble, as illustrated in Fig.\ \ref{fig:fig02}a. The comparison of $H$ with different grid resolutions at $\text{We}=227$ is shown in Fig.\ \ref{fig:fig02}a. We can see that when the grid resolution ${{D}_{0}}/\Delta x$ increases from 42 to 84, the maximum difference in $H$ after the droplet pinch-off is about 9.1$\%$ at 23.7 ms. However, when ${{D}_{0}}/\Delta x$ increases from 84 to 168, $H$ is very close, with a maximum difference of 2.2$\%$ at 23.7 ms, and thus the grid resolution of ${{D}_{0}}/\Delta x=84$ is sufficient for further simulations of the cavity deformation and the bubble pinch-off. Fig.\ \ref{fig:fig02}b shows the gas-liquid interface profiles at 20.5 ms. The bubble pinch-off could not be captured using the coarse grid (${{D}_{0}}/\Delta x=21$) at $\text{We}=227$. If the grid resolution is higher than ${{D}_{0}}/\Delta x=84$, the interface profiles were almost the same, and the bubble pinch-off could be captured well. Therefore, the grid resolution of ${{D}_{0}}/\Delta x=84$ is sufficient for further simulations of the cavity deformation and the bubble pinch-off. In addition, a grid independence study was also performed when the droplets impact at a higher We of 571 (the maximum Weber number considered in this study), and the results (Fig.\ \ref{fig:fig02}c) show that if the grid resolution is higher than ${{D}_{0}}/\Delta x=84$, the cavity depth $H$ is also independent of the grid resolution before 27 ms. After that is the formation of the Worthington jet, which is very fast, and there is still a slight deviation between the grids of ${{D}_{0}}/\Delta x=84$ and ${{D}_{0}}/\Delta x=168$. The Worthington jet is not our main concern in this study because if the phenomenon of the regular bubble does not occur, we only consider the time evolution of the cavity before the cavity achieves its maximum depth. Therefore, the grid resolution of ${{D}_{0}}/\Delta x=84$ is sufficient for the simulations at a higher Weber number. To check the grid independence at higher environmental pressure, a grid-independence study was also performed when the droplet impacts at $P^*=30$ (the maximum environmental pressure considered in this study), as shown in Fig.\ \ref{fig:fig02}d. In this scenario, the entrainment of a large bubble occurs soon (at about 3 ms) after the droplets impact at $\text{We}=279.9$. The results show that the cavity depth when the droplet impacts at a higher environmental pressure is very close in different grid resolutions. Therefore, a grid resolution of ${{D}_{0}}/\Delta x=84$ was used for further simulations of the cavity deformation and the bubble entrainment.

\section{Results and discussions}\label{sec:sec4}
\subsection{Model validation}\label{sec:sec41}
\begin{figure*}
  \centering
  \includegraphics[width=1.8\columnwidth]{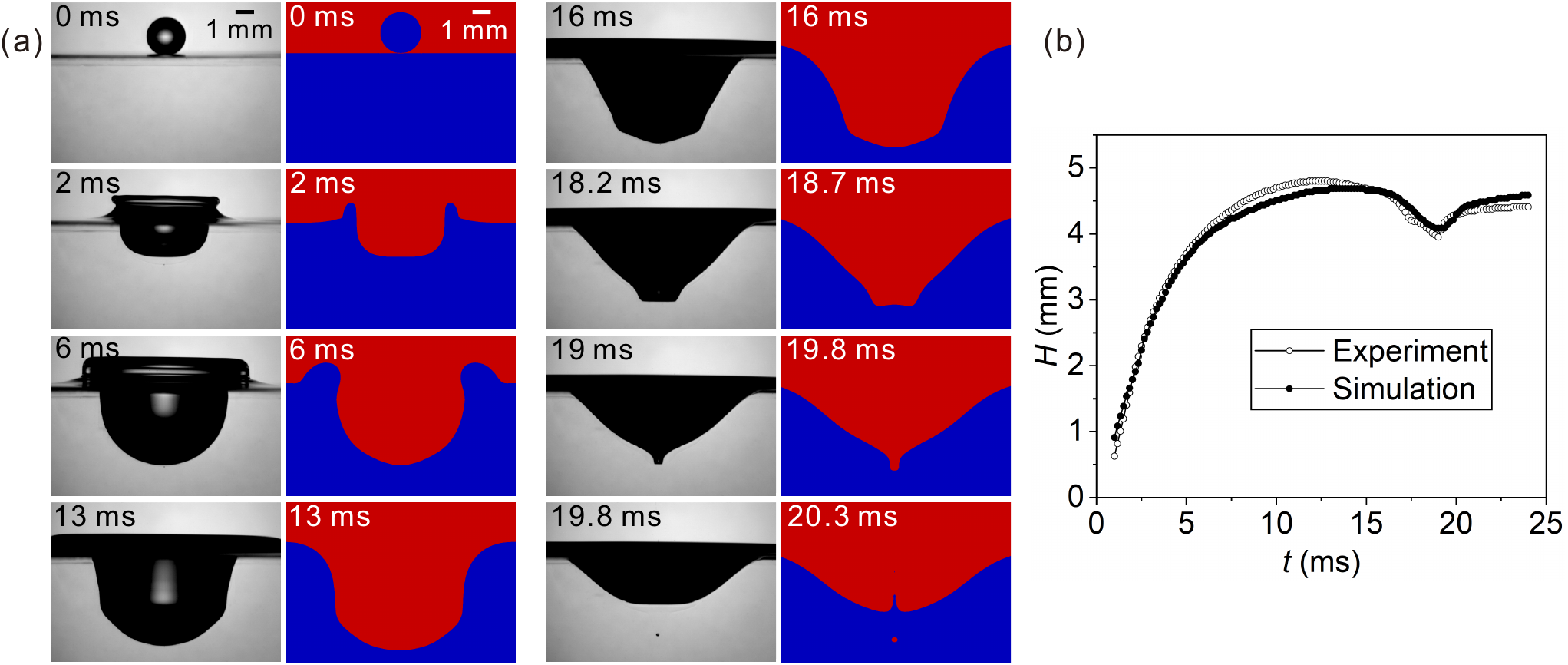}\\
  \caption{Comparisons between the numerical and experimental results after a droplet impacting a liquid pool. (a) Morphological variations (see Videos 1 and 2 in Supplementary Material \cite{SM}). (b) Cavity depth and the distance from the pool surface to the bottom of the regular bubble. $D_0=2.07$ mm and $\text{We}=227$. }\label{fig:fig03}
\end{figure*}

The accuracy of numerical simulation is validated against the droplet impact experimental data. The instant of the droplet impact is set as $t = 0$. The comparison of surface profiles between numerical simulation and experimental images is shown in Fig.\ \ref{fig:fig03}a. The cavity develops into a hemisphere and then retracts, and a regular bubble is pinched off in the end. At the very beginning stages of the impact in the experiment, a very small bubble is seen at the base of the cavity. The small bubble is from the rupture of the air film between the bottom of the droplet and the pool surface \cite{Tang2019InterfacialGas, Jian2020SplitAirDisk, Qi2020AirHeatedSubstrates, Tran2013AirEntrainmentLiquid, Lo2017AtomicallySmoothSubstrate, Langley2017UltraViscousDrops}. Since it is very small and happens at the very early stage of the impact process, its effect on the cavity deformation and the subsequent bubble entrapment is negligible. Since this study mainly focuses on the evolution of the air cavity and the subsequent bubble entrapment, we do not attempt to resolve the formation of the small bubble in the numerical simulation, which requires a grid resolution of sub-micrometer and is very computational demanding. The comparison shows good agreement between the simulation and the experiment in the time evolution of the free surface profile. There is a slight mismatch in time between the numerical and experimental results in the last three frames. One reason is that the error in time accumulates during the simulation of the cavity evolution which is much longer than the pinch-off of the regular bubble, and another reason is the one-frame uncertainty in determining the exact time of drop impact ($t = 0$) in the experimental sequences.

The cavity depth (before the bubble pinch-off) and the distance from the pool surface to the bottom of the regular bubble (after the bubble pinch-off) are used for the quantitative assessment of the numerical simulation. The comparisons between numerical and experimental results are shown in Fig.\ \ref{fig:fig03}b. The cavity depth increases as it expands and then decreases as it retracts. After 20 ms, \emph{H} increases further as the bubble moves downward in the liquid pool. The comparison shows good agreement between the simulation and the experiment, and indicates that the simulation can predict the cavity deformation and the bubble pinch-off.

\subsection{Effects of gravity}\label{sec:sec42}

\begin{figure*}
  \centering
  \includegraphics[width=1.2\columnwidth]{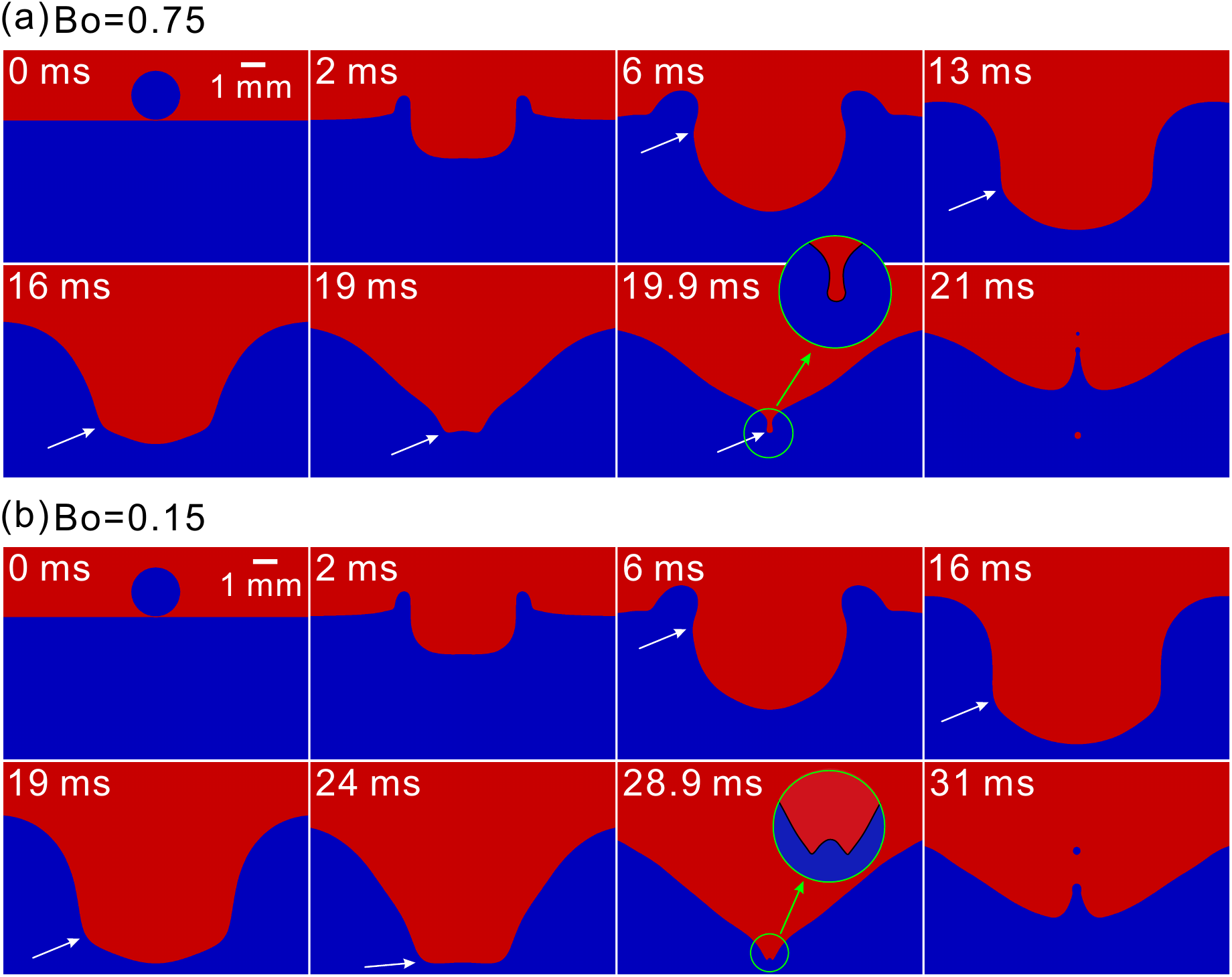}\\
  \caption{Cavity evolution after droplet impact at different gravities (see Videos 3 and 4 in Supplementary Material \cite{SM}). $D_0=2.1$ mm and $\text{We}=223.1$. The front of the capillary wave is highlighted by white arrows. The cylindrical air pocket is also highlighted by the magnified image at 19.9 ms in (a). }\label{fig:fig04}
\end{figure*}

\begin{figure*}
  \centering
  \includegraphics[width=1.25\columnwidth]{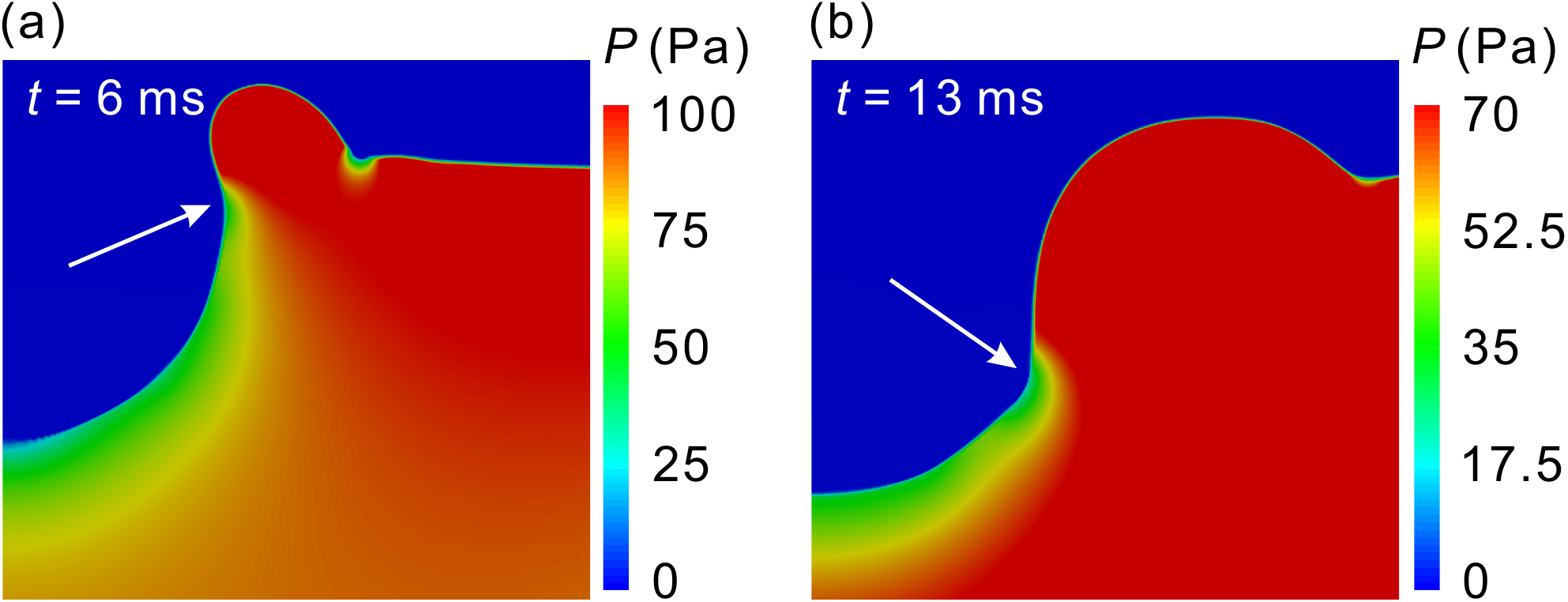}\\
  \caption{Pressure fields during the propagation of capillary wave. (a) $t = 6$ ms, (b) $t = 13$ ms. $\text{Bo}=0.75$, $D_0=2.1$ mm and $\text{We}=223.1$.}\label{fig:fig05}
\end{figure*}

To analyze the effect of the gravitational force on the cavity deformation and bubble pinch-off, we study the cavity behaviors by varying the gravity from 0 to 10 m/s$^2$ while fixing the other parameters. The evolution of the cavity at different Bond numbers is shown in Fig.\ \ref{fig:fig04}. A regular bubble is pinched off at earth's gravity ($\text{Bo}=0.75$, see Fig.\ \ref{fig:fig04}a). In contrast, at low gravity ($\text{Bo}=0.15$), the phenomenon of regular bubble pinch-off disappears (see Fig.\ \ref{fig:fig04}b).

The front of the capillary wave can be seen from the sharp shape on the surface of the cavity, which is highlighted by white arrows in Fig.\ \ref{fig:fig04}, and the origin of the capillary wave is due to the strong surface disturbance immediately after the initial contact of the impacting drop with the undisturbed liquid surface \cite{Morton2000FlowRegime, Berberovic2009LiuidLayer, Oguz1991CalculationUnderwaterNoise, Kassim2004DropCoalescenceInterface}. When a capillary wave is created, the sharp curvature of the interface at the front of the capillary wave induces a strong variation of the local pressure \cite{Berberovic2009LiuidLayer}. The pressure distribution (see Fig.\ \ref{fig:fig05}) shows that the front of the capillary wave separates the high-pressure region and the relatively low-pressure region. Therefore, the front of the capillary wave can be identified by the sharp curvature on the surface of the cavity. At earth's gravity, the capillary wave propagates downward along the cavity surface, as shown at 13 ms in Fig.\ \ref{fig:fig04}a. The capillary wave reaches the lowest point at the cavity bottom at 19.9 ms, producing a cylindrical air pocket (like a U shape) at the cavity bottom, as highlighted by the magnified image at 19.9 ms in Fig.\ \ref{fig:fig04}a. Then the mouth of the cylindrical air pocket closes, and a regular bubble is pinched off. In contrast, at low gravity ($\text{Bo}=0.15$), the interface at the cavity bottom alters its moving direction to the upward at 28.9 ms (see the magnified interface shape), and a jet is produced finally without regular bubble pinch-off.

\begin{figure*}
  \centering
  \includegraphics[scale=0.55]{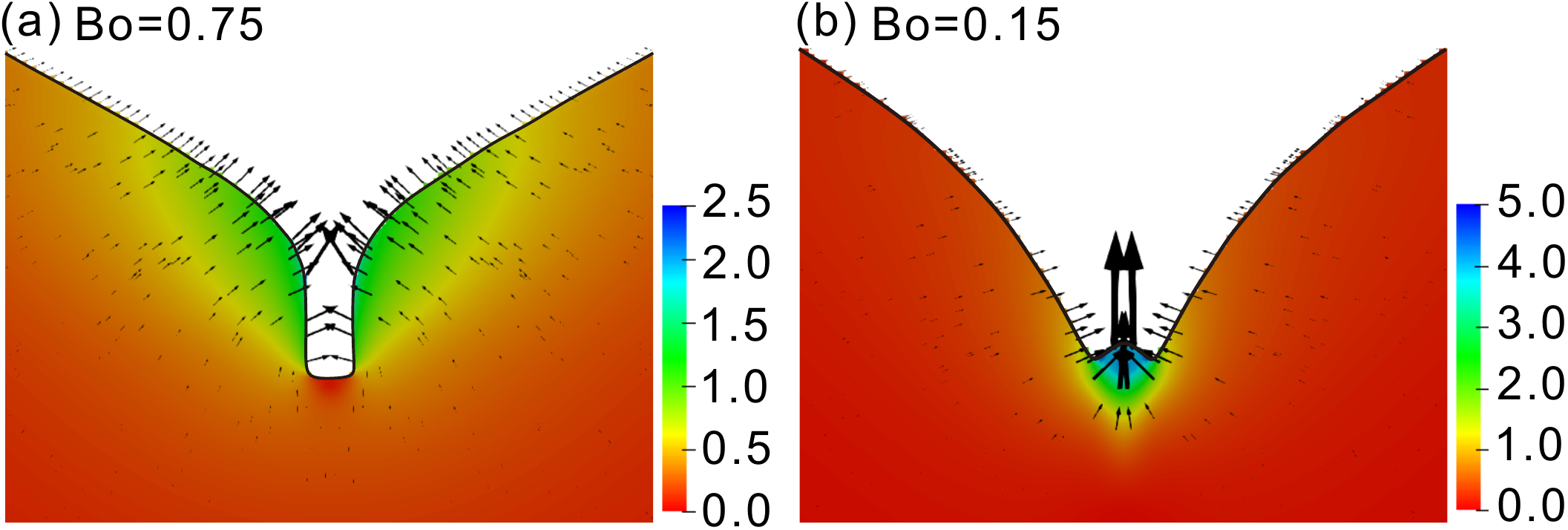}\\
  \caption{Comparison of the velocity fields at different gravities. (a) $\text{Bo}=0.75$, $t=19.9$ ms, with regular bubble pinch-off, (b) $\text{Bo}=0.15$, $t=28.9$ ms, no bubble pinch-off. $D_0=2.1$ mm and $\text{We}=223.1$.}\label{fig:fig06}
\end{figure*}

To further clarify the difference between the two scenarios, the velocity fields are compared in Fig.\ \ref{fig:fig06}. At normal gravity ($\text{Bo}=0.75$), the cavity surface moves inward, and the speed of the cavity surface is largest at the mouth of the air cylinder, and the speed of the interface at the bottom of the air cylinder is almost zero, as shown in Fig.\ \ref{fig:fig06}a. Thus, the mouth of the cylinder closes at high speed to pinch off a regular bubble. In contrast, at low gravity ($\text{Bo}=0.15$), the interface at the bottom of the cavity moves upward at high speed, which results in a high-speed jet and prevents the bubble pinch-off, as shown in Fig.\ \ref{fig:fig06}b.

\begin{figure*}
  \centering
  \includegraphics[scale=0.55]{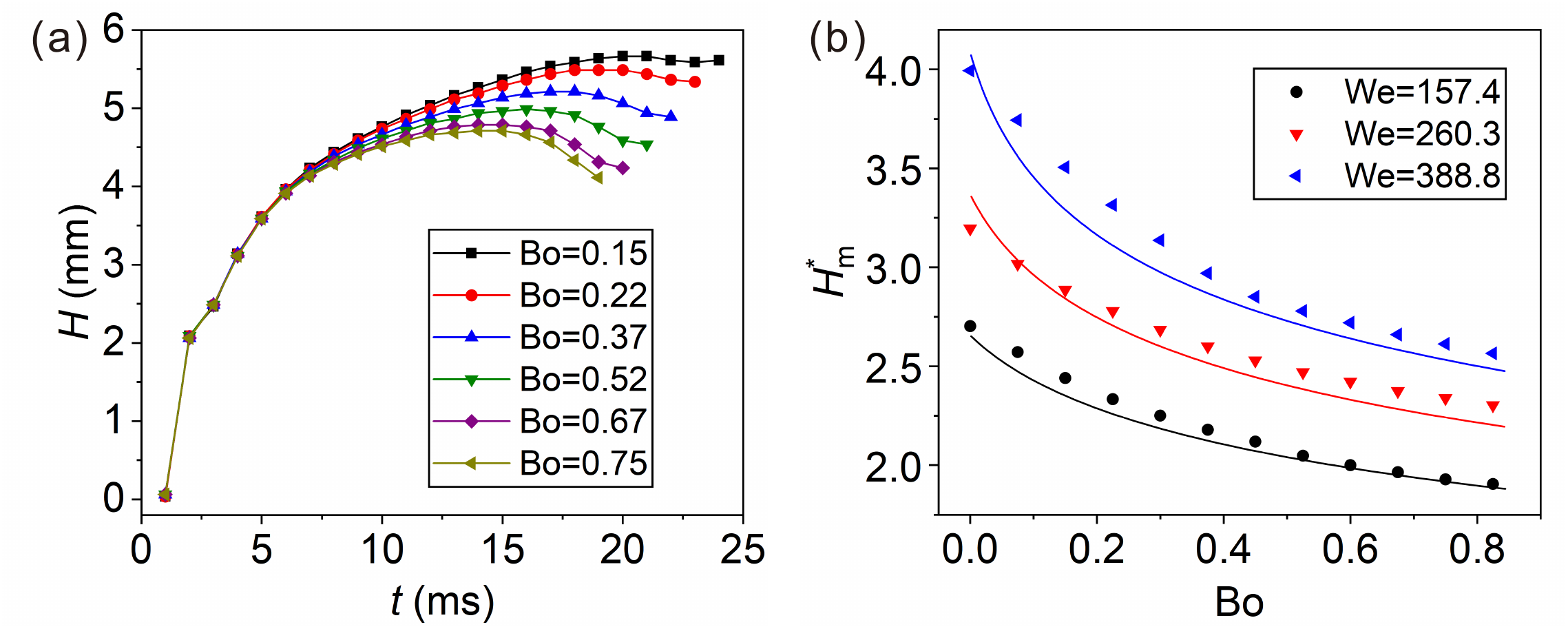}\\
  \caption{Effects of gravity on cavity deformation. (a) Time evolution of the cavity depth before the formation of the Worthington jet at different gravities. $D_0=2.1$ mm and $\text{We}=223.1$. (b) Maximum depth of the cavity $H_m^*$ at different gravities. The symbols are the results of numerical simulation, and the solid lines are theoretical results predicted by Eq.\ (\ref{eq:eq12}).}\label{fig:fig07}
\end{figure*}

The evolution of the cavity depth before the formation of the Worthington jet is shown in Fig.\ \ref{fig:fig07}a. Initially, the cavity depth $H$ increases rapidly due to the large inertia of the droplet fluid. As the droplet fluid sinks and the cavity expands, the growth of the cavity slows down, and the growth rate of the cavity depth decreases. After reaching a maximum cavity size, the cavity retracts due to the gravitational force. Therefore, the cavity depth decreases.
The maximum depth of the cavity ${{H}_{m}}$ is defined as the maximum depth measured from the initial pool surface to the bottom of the cavity along the axis of symmetry ($r=0$) before the formation of the Worthington jet. The time variation of the cavity depth in Fig.\ \ref{fig:fig07}a also shows that the maximum cavity depth is affected by gravitational force. With increasing the gravity, the maximum cavity depth decreases, and the time to reach the maximum cavity depth decreases. This can be explained by the conversion and conservation of energy. The droplet's initial energy ${{E}_{0}}$ includes kinetic and surface energy. The kinetic energy can be expressed as
\begin{equation}\label{eq:eq07}
  {{E}_{k0}}=\frac{1}{12}\pi \rho D_{0}^{3}U_{0}^{2}.
\end{equation}
The initial surface energy of the droplet is
\begin{equation}\label{eq:eq08}
  {{E}_{s0}}=\pi \sigma D_{0}^{2}.
\end{equation}

The droplet's initial kinetic energy and the surface energy are converted to the gravitational potential energy of the liquid crown and the surface energy of the cavity liquid. The increment in the gravitational potential energy is due to the liquid originally in the cavity position (before the droplet impact) moving to a higher position and forming the liquid crown (after the droplet impact). Therefore, the increment in the gravitational potential energy as the cavity expands to its maximum depth ${{H}_{m}}$ can be approximated as ${{E}_{p}}=mg{{H}_{m}}$, where \emph{m} is the mass of the elevated liquid and the elevated height of the liquid can be estimated by ${{H}_{m}}$. Since the cavity can be approximated by a hemisphere when it expands to the maximum depth \cite{Berberovic2009LiuidLayer, Bisighini2010CraterEvolution, Engel1966CraterDepth, Leng2001SplashSphere}, the mass of the elevated liquid can be expressed as $m=\rho V=2\pi \rho H_{m}^{3}/3$. Therefore, the increment in the gravitational potential energy of the liquid crown as the cavity expands to its maximum depth can be expressed as
\begin{equation}\label{eq:eq09}
  {{E}_{p}}=\frac{2}{3}\pi \rho gH_{m}^{4}.
\end{equation}
We ignore the kinetic energy of the liquid crown because the cavity can be assumed as quasi-stationary when it achieves its maximum depth. A precise calculation of the surface energy should consider the change in the surface area from the initial flat surface of the pool to the cavity and the crown, which is complex geometry and difficult to calculate directly. Since the cavity is almost hemispherical, we can simply estimate the scale of the variation of the surface area as $2\pi H_{m}^{2}$. Hence, the variation in the surface energy can be estimated as
\begin{equation}\label{eq:eq10}
  {{E}_{s}}=2\pi \sigma H_{m}^{2}.
\end{equation}
The droplet's initial energy is approximately the sum of the increment of the surface energy of the cavity and the gravitational potential energy of the liquid crown,
\begin{equation}\label{eq:eq11}
  {{E}_{k0}}+{{E}_{s}}_{0}={{E}_{p}}+{{E}_{s}}.
\end{equation}
By substituting Eqs.\ (\ref{eq:eq07})-(\ref{eq:eq10}) into Eq.\ (\ref{eq:eq11}), we can get
\begin{equation}\label{eq:eq12}
  \text{We} = 8\text{Bo}{{(H_{m}^{*})}^{4}}+24{{(H_{m}^{*})}^{2}}-12,
\end{equation}
where $H_{m}^{*}$ is the maximum dimensionless depth of the cavity and is defined as $H_{m}^{*}\equiv {{H}_{m}}/{{D}_{0}}$.

The maximum depth of the cavity at different gravities is shown in Fig.\ \ref{fig:fig07}b. The symbols are the results of the numerical simulation, and the lines are the theoretical results predicted by Eq.\ (\ref{eq:eq12}). The comparison shows reasonable agreement between the simulation and the theoretical prediction. The maximum depth of the cavity decreases with increasing $\text{Bo}$, and increases with increasing $\text{We}$. The reason is that the droplet with a larger $\text{We}$ has more inertia, which can be converted to more surface energy of the cavity, and thus produces a deeper cavity.

\begin{figure}
  \centering
  \includegraphics[scale=0.38]{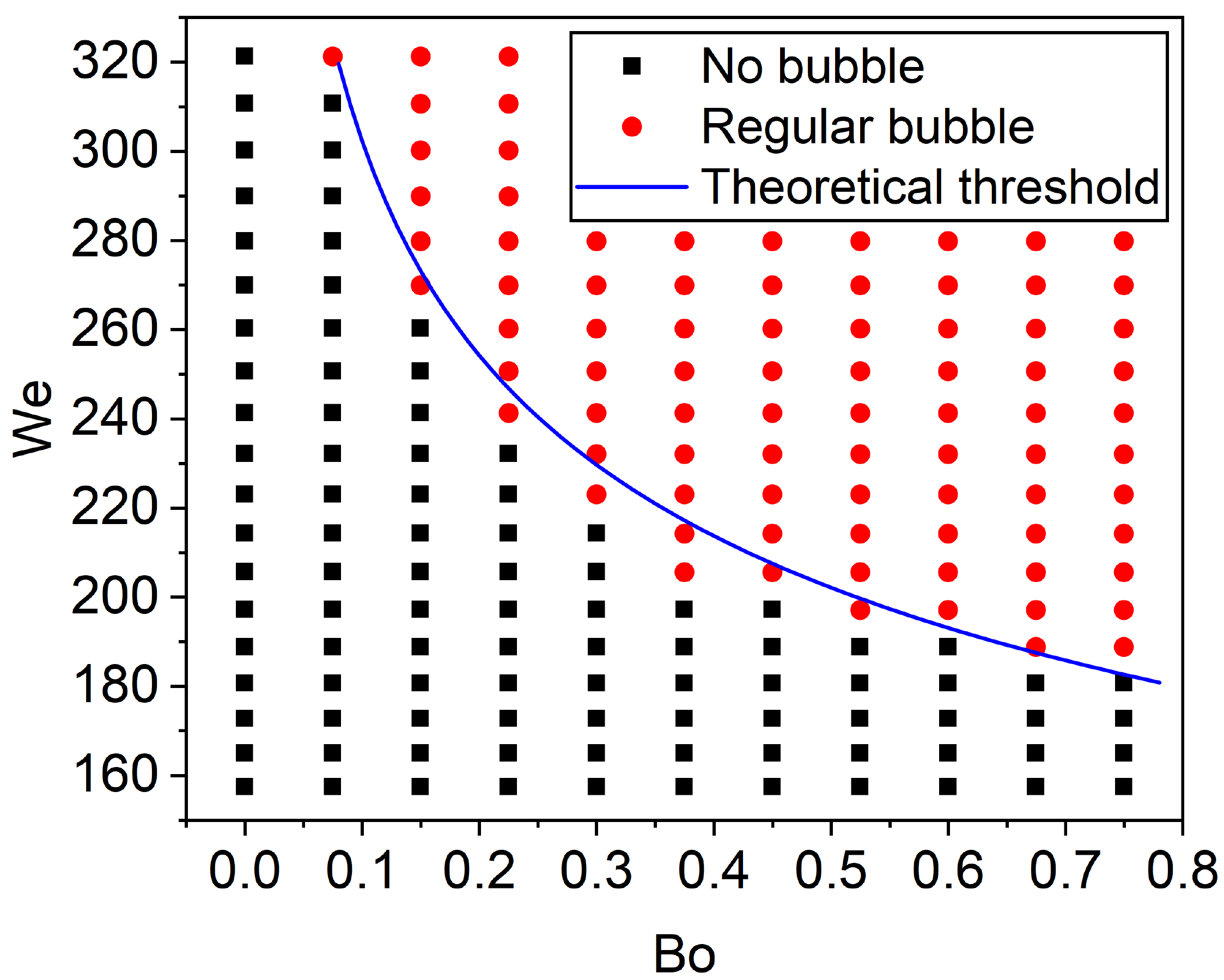}\\
  \caption{Regime map of regular bubble pinch-off after droplet impact at different gravities. The solid line is the theoretical threshold of the bubble pinch-off predicted by Eq.\ (\ref{eq:eq15}) with a fitting coefficient of 170. $D_0=2.1$ mm. }\label{fig:fig08}
\end{figure}

To determine the condition of regular bubble pinch-off, we produce a map of impact outcomes by varying the droplet speed $U_0$ and the gravitational acceleration $g$, as shown in Fig.\ \ref{fig:fig08}. Regular bubbles are produced when the inertia effect is strong (i.e., large We) and the gravitational force is large (i.e., large Bo). After the impact, capillary waves are formed near the impact point, and propagates inward and downward along the cavity surface. The pinch-off of a regular bubble occurs if the timescale for the cavity to achieve its maximum depth $t_m$ is smaller than the timescale of the capillary wave $t_w$ \cite{Morton2000FlowRegime, Oguz1990BubbleEntrainment}. The timescale for the cavity to achieve its maximum depth \cite{Oguz1990BubbleEntrainment} can be expressed as
\begin{equation}\label{eq:eq13}
  {{t}_{m}}\sim c_{{m}}^{-4/3}U_{0}^{1/3}{{D}_{0}},
\end{equation}
where ${{c}_{m}}\sim {{(g\sigma /\rho )}^{1/4}}$ is the minimum speed of the capillary-gravity surface wave. The timescale of capillary wave is
\begin{equation}\label{eq:eq14}
  {{t}_{w}}\sim {{\left[ (\sigma /\rho ){{k}^{3}} \right]}^{-1/2}},
\end{equation}
where $k$ is the wavenumber, $k\sim {{({{D}_{0}}\text{We})}^{-1}}$ \cite{Oguz1990BubbleEntrainment}. Therefore, the threshold of regular bubble pinch-off can be expressed as
\begin{equation}\label{eq:eq15}
  \text{We}\sim {{\text{Bo}}^{-1/4}}.
\end{equation}

The line in Fig.\ \ref{fig:fig08} is the theoretical threshold of the bubble pinch-off predicted by Eq.\ (\ref{eq:eq15}). The threshold from the numerical results agrees well with scaling model predicted by Eq.\ (\ref{eq:eq15}), confirming the accuracy of the above theoretical analysis. The analysis also shows that the main role of the gravitational force in the pinch-off of regular bubbles is to induce capillary-gravity surface waves. If the gravitation force is strong, the capillary-gravity waves propagate quickly to create a cavity of a hemispherical shape. Then the capillary wave propagates to the lowest point of the cavity, producing a cylinder of air. Finally, the cylindrical air pocket pinches off to produce a regular bubble. If the gravitational force increases (i.e., Bo increases), the capillary-gravity waves propagate faster, and the pinch-off of regular bubbles becomes easier. Therefore, the threshold Weber number of bubble pinch-off decreases as the Bond number increases.

\subsection{Effects of the environmental pressure}\label{sec:sec43}

\begin{figure*}
  \centering
  \includegraphics[width=1.6\columnwidth]{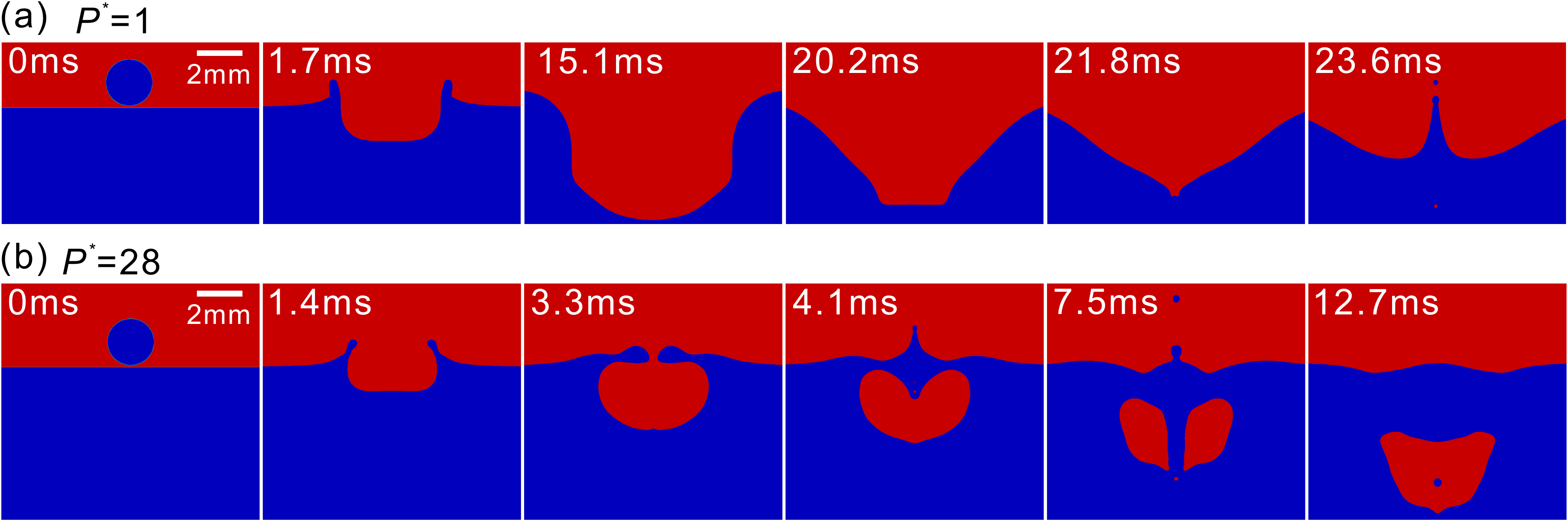}\\
  \caption{Cavity behaviors after the impact at different environmental pressures. (a) Pinch-off of a regular bubble at $P^*=1$; (b) Entrapment of a large bubble at $P^*=28$ (see Video 5 in Supplementary Material \cite{SM}). $D_0=2.1$ mm and $\text{We}=279.9$.}\label{fig:fig09}
\end{figure*}

Environmental pressure plays an essential role in the impact of droplets on a liquid pool \cite{Klyuzhin2010PersistingDropletsSurfaces, Marcotte2019EjectaCorollaSplashes}, and its effect on bubble entrapment is discussed in this section. The evolution of the cavity at different environmental pressures is compared in Fig.\ \ref{fig:fig09}. The relative pressure is used $P^*\equiv P/{{P}_{0}}$, where $P$ is the environmental pressure, and ${{P}_{0}}$ is the standard pressure, 1 bar. At standard environmental pressure $P^*=1$, the crown moves upward and also outward, as shown at 1.7 ms in Fig.\ \ref{fig:fig09}a. The capillary wave propagates inward along the cavity interface, and a cylindrical air pocket is produced at the bottom of the cavity at 21.8 ms, and the mouth of the cylindrical air pocket contracts to pinch off a regular bubble at 23.6 ms. In contrast, when the environmental pressure is high, the liquid crown moves upward and also inward, as shown at 1.4 ms in Fig.\ \ref{fig:fig09}b for $P^*=28$. The liquid crown becomes thicker as it moves inward. Then, the rim of the liquid crown collides, sealing a large bubble below the surface of the pool. Meanwhile, an upward jet and a downward jet are produced because of the inertia of the collision of the crown rim. Then, the downward jet punctures into the large bubble, which finally converges at the center. Meanwhile, the upward jet breaks up into secondary droplets owing to the jet's capillary instability.

\begin{figure*}
  \centering
  \includegraphics[width=1.65\columnwidth]{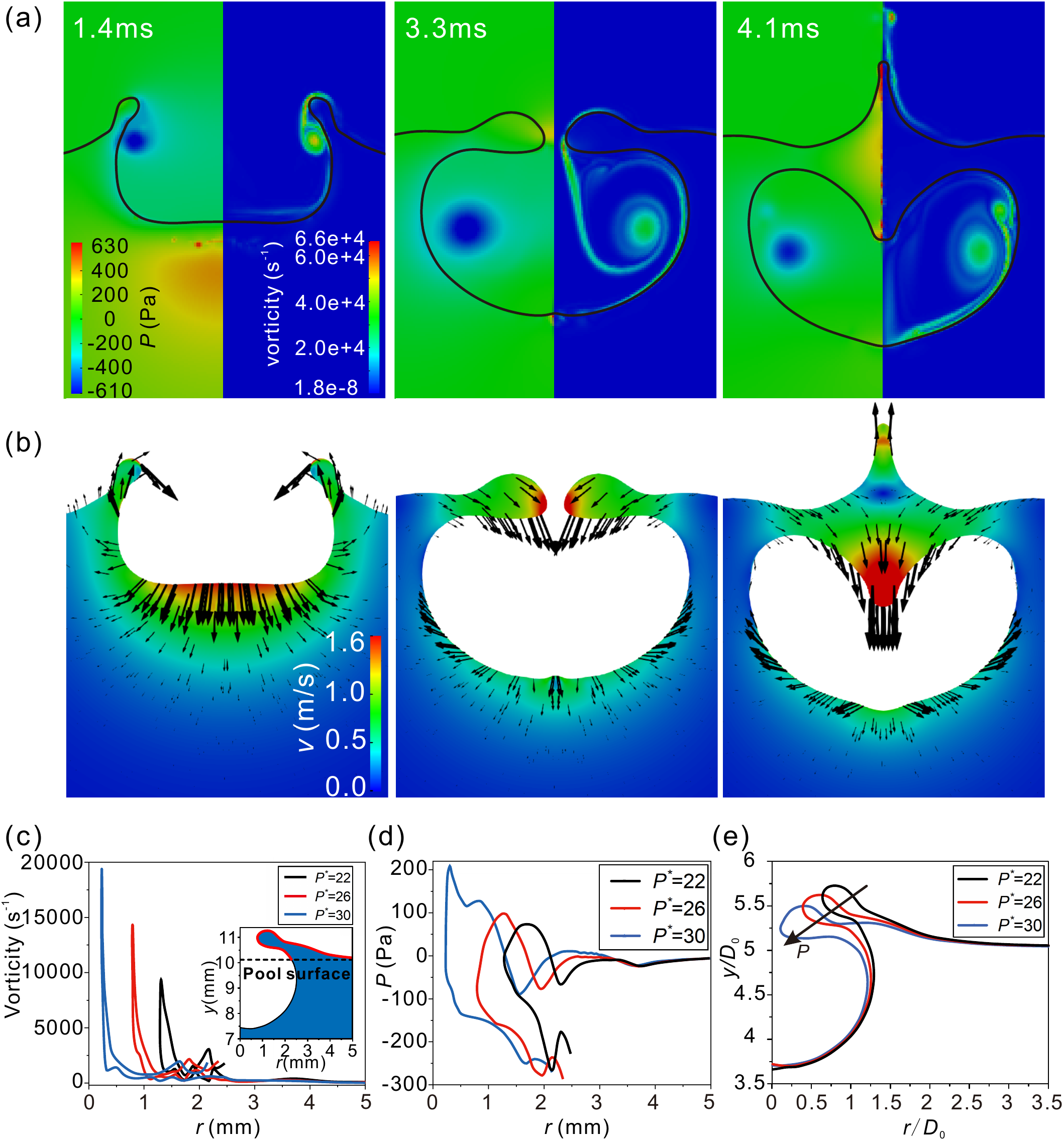}\\
  \caption{(a) Pressure fields (left) and vorticity fields (right) of large bubble entrapment. (b) Velocity fields of large bubble entrapment. The relative pressure is $P^*=28$. (c) Vorticity along the surface of liquid crown. The surface of the crown is highlighted by a red line in the inset. (d) Pressure along the surface of liquid crown. (e) Shape of the liquid crown rim. $t=3$ ms. $D_0=2.1$ mm and $\text{We}=279.9$. }\label{fig:fig10}
\end{figure*}

To analyze the mechanism of large bubble entrapment, we consider the vorticity field and the pressure field. Upon the impact, strong vorticity is induced at the interface. Then the vorticity detaches from the interface and forms a vortex ring. The pressure, vorticity, and velocity fields are shown in Figs.\ \ref{fig:fig10}a and \ref{fig:fig10}b. A pressure difference is found in the air phase between the two sides of the liquid crown, and the vorticity detaches from the crown surface to form a vortex ring, as shown at 1.4 ms in Fig.\ \ref{fig:fig10}a. The variation of the vorticity and the pressure along the surface of the crown at a typical instant are plotted in Figs.\ \ref{fig:fig10}c and \ref{fig:fig10}d, respectively. As the environmental pressure increase, the vorticity becomes stronger, and the pressure difference between the two sides of the crown rim becomes larger. The stronger vortex and the larger pressure difference push the rim to bend inward and downward (see the velocity field in Fig.\ \ref{fig:fig10}b and the shape of the crown rim in Fig.\ \ref{fig:fig10}e). The crown rims finally contact and entrap a large bubble.

\begin{figure}
  \centering
  \includegraphics[scale=0.35]{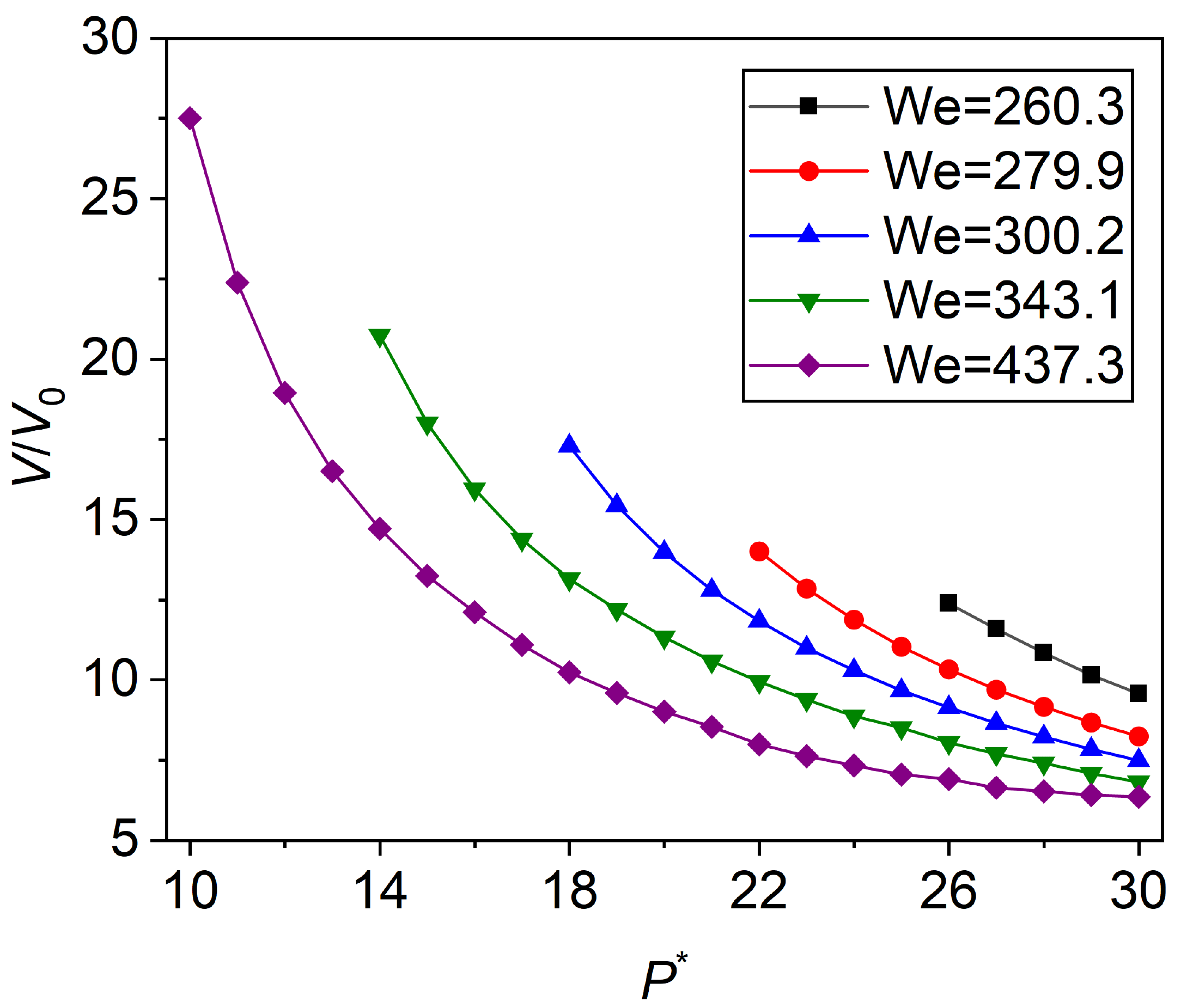}\\
  \caption{The normalized volumes of large bubbles entrapped after droplet impact at different environmental pressures. $V_0$ is the volume of the droplet. $D_0=2.1$ mm.}\label{fig:fig11}
\end{figure}

To characterize the entrapped large bubbles, we plot the volumes of large bubbles at different environmental pressures in Fig.\ \ref{fig:fig11}. The bubble can be much larger than the initial droplet size. For example, at $P^*=10$ and $\text{We}=437.3$, the volume of the large bubble is almost 27.5 times the initial droplet volume. The volume of the large bubble decreases with increasing environmental pressure. As environmental pressure increases, vorticity becomes stronger, and the pressure difference between the two sides of the crown rim becomes larger. The stronger vortex and the larger pressure difference push the crown to bend inward and downward. The bending degree of the crown increases with the environmental pressure (see Fig.\ \ref{fig:fig10}e). Thus, the volume of the entrapped large bubble decreases as we increase the environmental pressure. The maximum volume of the large bubble occurs at the left end of each curve for a fixed $\text{We}$, as shown in Fig.\ \ref{fig:fig11}. If the environmental pressure decreases further, the transition from large bubble to regular bubble occurs. The threshold for the transition will be discussed later in this section.

\begin{figure*}
  \centering
  \includegraphics[width=1.5\columnwidth]{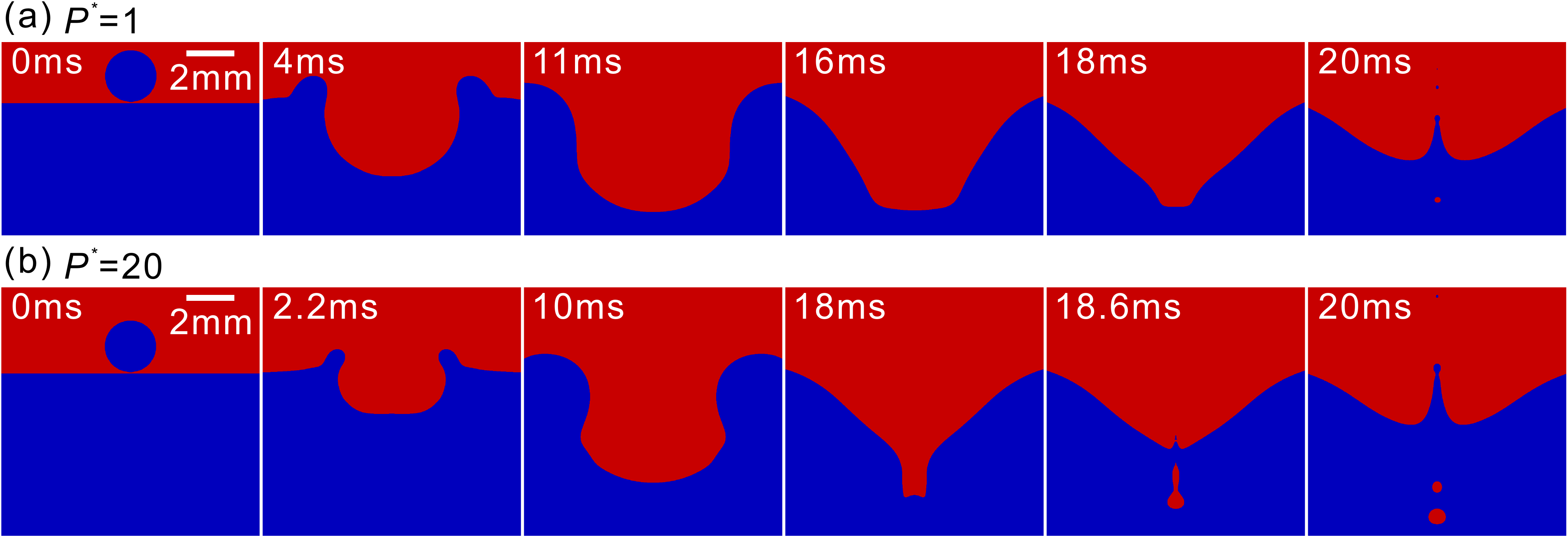}\\
  \caption{Evolution of the cavity after droplet impact: (a) Entrapment of a single bubble at $P^*=1$. (b) Entrapment of two bubbles at $P^*=20$ (see Video 6 in Supplementary Material \cite{SM}). $D_0=2.1$ mm, and $\text{We}= 188.9$.}\label{fig:fig12}
\end{figure*}

At high environmental pressure, it is also possible to entrap two bubbles after the impact. The evolution of the cavity for two-bubble entrapment in a typical case is compared with that for single bubble entrapment in Fig.\ \ref{fig:fig12}. At the standard environmental pressure, a regular bubble is pinched off at 20 ms, as shown in Fig.\ \ref{fig:fig12}a. In contrast, when the relative pressure is 20, two bubbles were pinched off as shown at 20 ms in Fig.\ \ref{fig:fig12}b. As the capillary wave propagates inward along the interface of the cavity, a cylinder of air is formed at 18 ms. By comparing with the air cylinder at $P^*=1$, we can see that the air cylinder at $P^*=20$ is much longer than that at $P^*=1$ (see the size of the air cylinder at 18 ms in Fig.\ \ref{fig:fig12}a and \ref{fig:fig12}b). Therefore, when the mouth of the air cylinder closes, the air bubble produced from the cylinder is very long. Consequently, the long bubble further pinches off into two bubbles by collapsing at the waist, as shown at 18.6 ms and 20 ms in Fig.\ \ref{fig:fig12}b.

\begin{figure}
  \centering
  \includegraphics[width=\columnwidth]{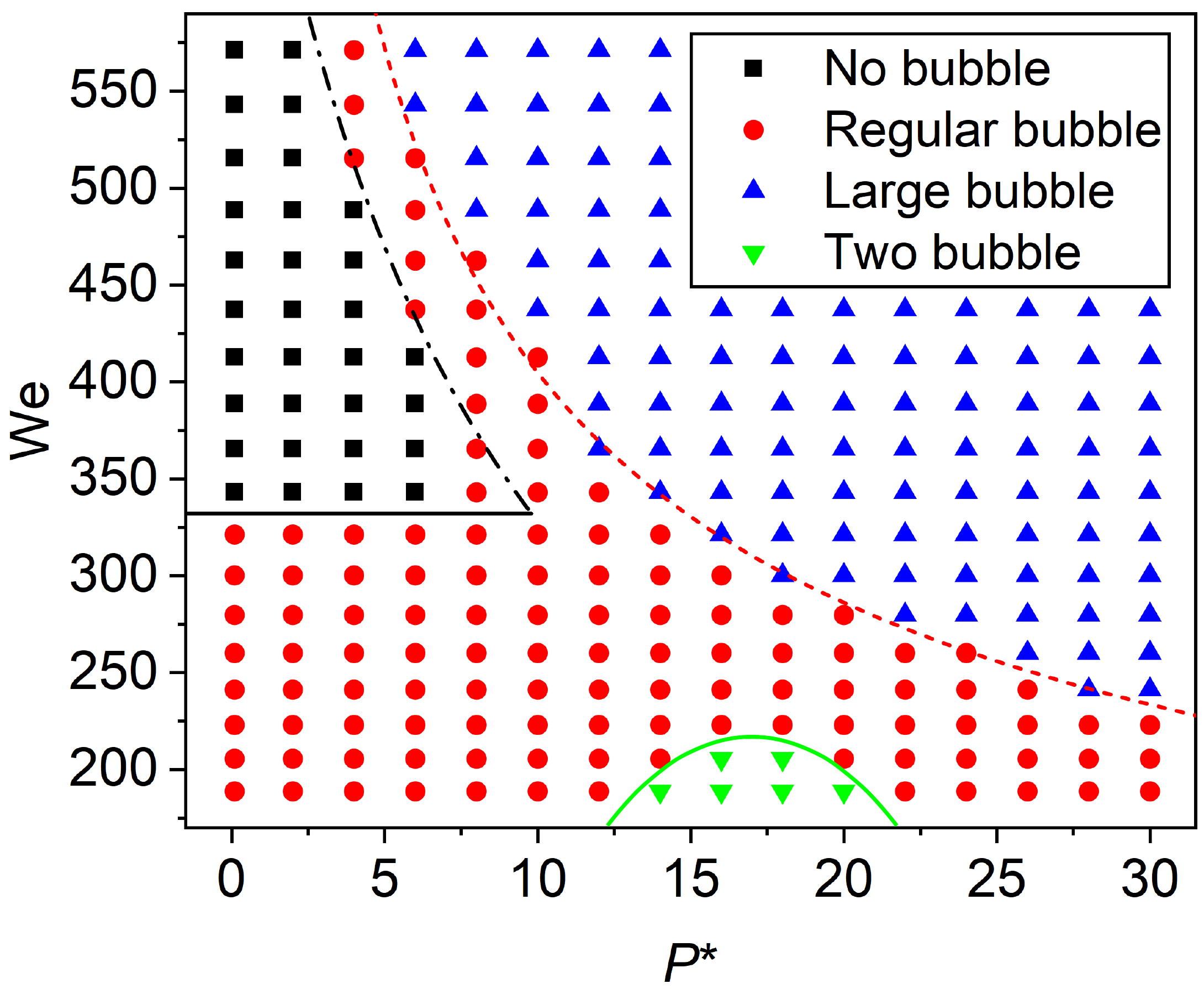}\\
  \caption{Regime map of bubble entrapment after droplets impact a liquid pool at different environmental pressures. The red dashed line is the threshold for the transition from regular bubble to large bubble entrapment predicted by Eq.\ (\ref{eq:eq16}) with a fitting coefficient of 1280 and $b = -0.5$. The black dash-dotted line is the threshold for the transition from no bubble to regular bubble pinch-off predicted by Eq.\ (\ref{eq:eq17}) with a fitting coefficient of 1180 and $n = -0.1$. The solid lines are just guides to the eye.}\label{fig:fig13}
\end{figure}

To determine the conditions of different regimes of bubble entrapment, we produce a regime map of the impact outcomes by varying the droplet speed and the environmental pressure. Large bubble entrapment occurs under a wide range of environmental pressure, as shown in Fig.\ \ref{fig:fig13}. A threshold for the transition from regular bubble pinch-off to large bubble entrapment is found, which is indicated by the red dashed line. Pan et al. \cite{Pan2010SurfaceProperties} studied the effect of the liquid viscosity on droplet impact and found that the threshold for the large bubble occurrence was almost not affected by the liquid viscosity. Therefore, the pressure threshold \emph{P} for the large bubbles occurrence at different environmental pressures can be expressed as a function of the liquid density $\rho$, the droplet diameter $D_0$, and the droplet speed $U_0$, i.e., $P=f(\rho ,\sigma ,{{D}_{0}},{{U}_{0}})$. Xu et al. \cite{Xu2017BurningEthanolSurface} studied the threshold for the large bubble occurrence based on dimensional analysis and suggested a power function of $f(\text{We})$, i.e., $f({\rm{We}}) \sim {\rm{W}}{{\rm{e}}^a}$, where \emph{a} is constant. Substituting it into $P=\rho U_{0}^{2}f(\text{We})$, the threshold for large bubble occurrence at different environmental pressures can be expressed as
\begin{equation}\label{eq:eq16}
  \text{We}\sim {(\sigma /{D_0}{P_0})^{ - b}}{P^*}^b,
\end{equation}
where $b\equiv 1/(a+1)$ is constant for this study, which can be obtained by fitting. As shown in Fig.\ \ref{fig:fig13}, the scaling agrees well with the numerical simulation.

Figure 13 also shows the transition threshold from no bubble to the regular bubble pinch-off.
%As shown in Fig.\ \ref{fig:fig09}, the threshold of Weber number decreases with increasing environmental pressure. The black line indicates the transition threshold from no bubble to the regular bubble pinch-off.
At $\text{We}>320$, there is only a narrow range of pressure for regular bubble pinch-off. As mentioned above, the pinch-off of a regular bubble occurs if the timescale for the cavity to achieve its maximum depth $t_m$ is smaller than the timescale of the capillary wave $t_w$. Chen et al. \cite{Chen2006PartialCoalescenceInterface} found that the density of the ambient fluid played an important role in capillary wave propagation when droplets impact a liquid surface. In our previous study \cite{Xu2022CrownRupturePressure}, we found that the environmental pressure mainly affected the density of the ambient gas during the droplet impact. We introduce the effect of the environmental pressure on the capillary wave propagation, and the minimum speed of the capillary-gravity surface wave can be expressed as ${{c}_{m}}\sim {{(g\sigma /\rho ')}^{1/4}}$, where $\rho '$ is the effective density $\rho '=\rho -m{{\rho }_{g}}$ with ${{\rho }_{g}}$ is the density of the ambient gas and \emph{m} is a constant coefficient. From Eq.\ (\ref{eq:eq14}), the timescale of the capillary wave is ${{t}_{w}}\sim {{[(\sigma /\rho '){{k}^{3}}]}^{-1/2}}$. Therefore, the threshold of regular bubble pinch-off can be obtained when the timescale for the cavity to achieve its maximum depth is equal to the timescale of the capillary wave: $\text{We}\sim \text{B}{{\text{o}}^{-1/4}}{{\left[ {{(\rho -m{{\rho }_{g}})}^{5}}/{{\rho }^{5}} \right]}^{-1/4}}$. Since the air density is proportional to the environmental pressure, ${{P}^{\text{*}}}=P/{{P}_{0}}={{\rho }_{g}}/{{\rho }_{0}}$, after rearrangement, the threshold for the transition from no bubble to the regular bubble pinch-off can be expressed in dimensionless form as
\begin{equation}\label{eq:eq17}
  \text{We}\sim \text{B}{{\text{o}}^{-1/4}}{{(1-n{{P}^{\text{*}}})}^{-5/4}},
\end{equation}
where $n\equiv m{{\rho }_{0}}/\rho $ is a constant coefficient, which can be obtained by fitting. As shown in Fig.\ \ref{fig:fig13}, the scaling agrees well with the numerical simulation.

 However, at $220<\text{We}<320$, as the environmental pressure increases, the process changes from regular bubble pinch-off to large bubble entrapment. When the Weber number is less than 320 at low environmental pressure, the timescale for the cavity to achieve its maximum depth is all smaller than the timescale of the capillary wave, ${{t}_{m}}<{{t}_{w}}$, so the regular bubble pinch-off occurs. For example, when $\text{We}=279.9$ at the standard environmental pressure (see Fig.\ \ref{fig:fig09}a), the time for the cavity to reach its maximum depth ${{t}_{m}}$ is about 15.1 ms, and the time of the capillary wave ${{t}_{w}}$ is about 21.8 ms (${{t}_{m}}<{{t}_{w}}$), thus the regular bubble pinch-off occurs. The phenomenon of two-bubble pinch-off occurs at low Weber number and moderate environmental pressure, as shown in Fig.\ \ref{fig:fig13}. When the environmental pressure is moderate, the liquid crown moves upward, and then sinks into the pool owing to the gravity and the surface tension force because the vorticity and the pressure difference are not enough to push the crowns to move inward to contact. As the capillary wave propagates inward along the cavity interface, a long cylinder of air is formed. Finally, two bubbles are produced when the mouth of the long cylinder of air closes and the waist pinches off.

\section{Conclusions}\label{sec:sec5}

In this study, the cavity deformation and bubble entrapment after the impact of droplets on a liquid pool are studied by combined experimental measurements and numerical simulations. The numerical simulations are validated against experimental results by comparing the evolution of the cavity. The regular bubble pinch-off is the result of the capillary wave propagating downward along the surface of the cavity and merging at the bottom of the cavity. The large bubble entrapment is the result of the liquid crowns merging at the mouth of the cavity. Gravity and environmental pressure play important roles in cavity deformation and bubble entrapment. The maximum depth of the cavity decreases as the gravitational effect becomes stronger. A regime map of bubble entrapment is obtained at different gravities. The phenomenon of regular bubble pinch-off may disappear as the gravity decreases. As the environmental pressure is increased, the regular bubble pinch-off is transformed into the large bubble entrapment. The process of the large bubble entrapment is controlled by the vortex ring at the gas-liquid interface and the pressure difference between the two sides of the liquid crown. The volume of the large bubble entrapped decreases with increasing the environmental pressure. A regime map of bubble entrapment is obtained at different environmental pressures. When the environmental pressure is high and the Weber number is large, the strong vorticity at the gas-liquid interface and the large pressure difference between the two sides of the rim make the crown rim collide to entrap a large bubble. When the environmental pressure is low, a regular bubble is pinched off at the bottom of the cavity due to the propagation of the capillary waves. Finally, the thresholds for the transitions from no bubble to regular bubble and to large bubble regimes are analyzed.

This paper reports an experimental and numerical study of a droplet impacting a liquid pool. There are still many open questions for the impact process, such as the subsequent dynamics of the entrapped bubble and the dynamics of the liquid jet produced after bubble entrapment. This study of the impact of droplets on a liquid pool will not only deepen our understanding of the impact dynamics but also will be useful for the optimization of the relevant applications.

\section*{Acknowledgements}
This work is supported by the National Natural Science Foundation of China (Grant Nos.\ 51676137, 51921004).

\section*{Conflict of interest}
The authors have no conflicts to disclose.

\section*{Data Availability Statement}
The data that support the findings of this study are available from the corresponding author upon reasonable request.
\bibliography{dropletImpactBubble}
\end{document}